\documentstyle[aps,epsfig,floats,preprint]{revtex}
\begin{document}
\draft
\tightenlines

\title{Determination of $\alpha_s$ from Gross-Llewellyn Smith
sum rule by accounting for infrared renormalon\footnote{
\footnotesize Partly based on invited talk given by G.C. in WG3 of NuFact02, 
July 1-6, 2002, Imperial College, London}}

\author{Carlos Contreras$^1$, Gorazd Cveti\v c$^1$, Kwang Sik Jeong$^2$ 
and  Taekoon Lee$^2$}
\address{$^1$Dept.~of Physics,
Universidad T\'ecnica Federico Santa Mar\'{\i}a,
Valpara\'{\i}so, Chile\\
$^2$Dept.~of Physics, Korea Advanced Institute of Science and Technology,
Daejon, Korea}
\maketitle

\begin{abstract}
We recapitulate the method which resums the truncated 
perturbation series of a physical observable in a way 
which takes into account the structure of the leading infrared 
renormalon. We apply the method to the Gross-Llewellyn Smith 
(GLS) sum rule. By confronting the obtained result with the 
experimentally extracted GLS value, we determine the value 
of the QCD coupling parameter which turns out to agree with 
the present world average.
\end{abstract}

\section*{}
The infrared (IR) renormalons are singularities
at positive values of the Borel variable $b$
in the Borel transforms of QCD physical observables.
These renormalons represent the contributions of 
loop integrations in the low energy regimes 
\cite{Mueller:vh}. They represent obstacles for and 
lead to ambiguities in the Borel integration. This
integration then gives discontinuities (cuts) of the observables
in the complex coupling parameter plane.
It is thus important to eliminate these unphysical cuts
from the Borel integration. 
This approach was proposed
in \cite{Lee:2001ws} and applied to
the GLS sum rule in \cite{Contreras:2002kf}. Here
we recapitulate the method and the results of
\cite{Contreras:2002kf}, and emphasize some
physical aspects.
The GLS sum rule is the quantity
\begin{eqnarray}
M_3(Q^2) \equiv 3 \left(1 - \Delta(Q^2) \right)
& =&  \int_0^1~dx~F_3(x;Q^2) \left[ 1
+ {\cal O} \left( \frac{m_N^2}{Q^2} \right) \right]
\ ,
\label{GLSdef}
\end{eqnarray}
where $F_3 = (F_3^{\nu p} + F_3^{\nu n})/2$ is
the nonsinglet structure function in the 
$\nu N$ DIS, $m_N$ is the nucleon mass.
Theoretical knowledge about the GLS
$\Delta(Q^2)$ is the following. The truncated
perturbation series (TPS) up to $\sim\!a^3$ 
[$a(Q) \equiv \alpha_s(Q)/\pi$] is known \cite{GLZN}
\begin{eqnarray}
\Delta(Q^2) &\sim& a (1 + w_1 a + w_2 a^2) \ ,
\label{GLSTPS}
\end{eqnarray}
as is also the structure of the first IR renormalon singularity
of its Borel transform
\begin{eqnarray}
{\rm BT}(b) & = & \frac{C}{ (1 - b)^{1 + \nu} } \left[
1 + \kappa_1 (1 - b) + \cdots \right] \ .
\label{IR1}
\end{eqnarray}
Here $\nu \!=\!(\beta_1/\beta_0\!-\!\gamma_2)/\beta_0$,
where $\beta_0$ and $\beta_1$ are the one- and two-loop
QCD beta function coefficients and $\gamma_2$ is the
one-loop coefficient of the anomalous dimension of the
twist-four ($d\!=\!2$) operator appearing in the Operator
Product Expansion (OPE) for $\Delta(Q^2)$. When the number 
of active quark flavors is $n_f\!=\!3$, we have $\beta_0\!=\!9/4$,
$\beta_1\!=\!4$, and $\gamma_2\!=\!8/9$ (the latter value 
obtained in \cite{Shuryak:1981kj}), giving
$\nu = 32/81$. The TPS of ${\rm BT}(b)$ in powers of
$b$ is known to $\sim\!b^2$, as a consequence of (\ref{GLSTPS}). 
A consequence of the IR singularity (\ref{IR1}) 
at $b \geq 1$ is that the Borel integral $\Delta_{\rm BI}(z)$, 
which for the complex coupling $z\!=\!|z| \exp(i \phi)$ is defined as 
\begin{equation}
\!\!\!\!\!\!\!\!\!\!\!\!\!\!\!\!
\Delta_{\rm BI}(z) = \frac{1}{\beta_0}~
\int_{0}^{+ \infty \exp(i \phi)}
~db~\exp \left[- \frac{b}{\beta_0 z} \right] {\rm BT}(b) \ , \qquad
(b = |b| e^{i \phi}) \ ,
\label{Bintegraldef}
\end{equation}
has a discontinuity (cut)
at the positive real axis $z\!=\!a(Q) \geq 0$. Namely,
for $z$ near the positive real axis, it can be shown from
(\ref{Bintegraldef}) that
\begin{equation}
\Delta_{\rm BI}(z) {\big |}_{z = a(Q) \pm i \varepsilon}
= \frac{1}{\beta_0}
~\int_{\pm i \varepsilon}^{\infty \pm i \varepsilon} 
~db~\exp \left[- \frac{b}{\beta_0 a(Q)} \right] {\rm BT}(b) \ .
\label{Bintegral}
\end{equation}
This quantity is not even real; the discontinuity shows
up in its imaginary part
\begin{eqnarray}
{\rm Im} \Delta_{\rm BI}(a(Q)\!\pm\!i \varepsilon) =
\frac{1}{\beta_0} {\rm Im}
\int_{\pm i \varepsilon}^{\infty\!\pm\!i \varepsilon} 
db \exp \left[- \frac{b}{\beta_0 a(Q)} \right]
\frac{C}{(1\!-\!b)^{1\!+\!\nu}} \left[ 1\!+\!\kappa_1 (1\!-\!b)\!+\!
\cdots \right]
\nonumber\\
\!\!\!\!\!\!\!\!\!\!\!\!\!\!\!\!
= \mp C {\beta_0}^{-1 - \nu} \left[ e^{-1/(\beta_0 a(Q))} \right]
a(Q)^{-\nu} \Gamma(-\nu) \sin( \pi \nu) \left[1 + \kappa_1 \beta_0 a \nu
+ \cdots \right] .
\label{ImDBI}
\end{eqnarray}
Because $(-a\!\mp\!i \varepsilon)^{-\nu}\!=\!
a^{-\nu} [ \cos(\pi \nu)\!\pm\!i \sin(\pi \nu)]$,
(\ref{ImDBI}) leads to
\begin{equation}
\Delta_{\rm BI}(z) =  - \frac{C \Gamma(-\nu)}{\beta_0^{1+\nu}}
\left[ e^{-1/(\beta_0 a(Q))} \right] \left[ (- z)^{-\nu}\!
+\!\kappa_1 (-\nu) \beta_0 (- z)^{-\nu +1}\!+\!\cdots \right] 
+ {\tilde \Delta}_{\rm BI}(z),
\label{DBI}
\end{equation} 
where ${\tilde \Delta}_{\rm BI}(z)$ is a function without
cuts in the complex $z$-plane since the first
expression on the right-hand side absorbs
the renormalon-induced cut at $z \geq 0$.
This expression, when $z\!=\!a(Q)$ is real, has the $Q$-dependence 
$Q^{-2} \alpha_s(Q)^{\gamma_2/\beta_0} [ 1\!+\!{\cal O}(\alpha_s) ]$, 
the same as the twist-four term in the OPE for the GLS sum rule
\cite{Shuryak:1981kj}. The Im part of this
expression, i.e.~expression (\ref{ImDBI}), can be identified 
as the Im contribution from the leading IR renormalon (\ref{IR1}).
The  premise of the method is 
that the full first expression on 
the right-hand side of (\ref{DBI}), i.e., including its real
part, represents the unphysical cut contribution
which is to be eliminated.
This leads to the final result for the resummed GLS value
\begin{equation}
\Delta(a(Q)) = {\tilde \Delta}_{\rm BI}(a(Q)) =
\left[ {\rm Re} \mp \cot(\pi \nu) {\rm Im} \right]
\int_{\pm i \varepsilon}^{\infty\!\pm\!i \varepsilon} 
db \exp \left[- \frac{b}{\beta_0 a(Q)} \right]
\frac{R(b)}{(1 - b)^{1 + \nu}} .
\label{res1}
\end{equation}
In the Borel integration here, the exactly known IR renormalon
singularity has been factored out explicitly. Function
$R(b)\!=\!(1 - b)^{1 +\nu} {\rm BT}(b)$, 
whose TPS up to $\sim\!b^2$ is known exactly due
to (\ref{GLSTPS}), has much weaker
singularity at $b\!=\!1$ than ${\rm BT}(b)$.
The function $R(b)$ has, in principle, singularities
from other renormalons, at $b=-1, \pm 2, \ldots$
To accelerate the convergence in
the TPS for $R$ further, we introduce in integral
(\ref{res1}) the change of variables $b \mapsto w$
according to the conformal mapping
\begin{eqnarray}
b \mapsto w(b) &=& \left( \sqrt{1 + b} - \sqrt{1 - b/2}  \right) 
\left( {\sqrt{1 + b} + \sqrt{1 - b/2} } \right)^{-1} \ ,
\label{ct}
\end{eqnarray}
which maps all these renormalon locations onto the unit circle
$|w|=1$, except for the leading IR renormalon
$w(b\!=\!1) = 1/3$ whose structure has been factored out.
Similar applications of conformal mappings have
been applied in other methods in Refs.~\cite{ct1,ct2}.
Further, expansion of $R(b(w))$
in powers of $w$ suggests that the ${\rm N}^3 {\rm LO}$
coefficient (at $a^4$) in the GLS TPS (\ref{GLSTPS}),
in ${\overline {\rm MS}}$
and with renormalization scale $\mu\!=\!Q$, is 
$w_3\!=\!158 \pm 30$, somewhat higher than
the TPS PMS prediction $w_3 \approx 130$\ \cite{KS}.
Using $w_3\!=\!158 \pm 30$, we use
for $R(b(w))$ in (\ref{res1}) the TPS up to $\sim\!w^4$.

Expression (\ref{res1}) represents the massless
part of the GLS sum rule. For $Q^2 \approx 2$-$3 \ {\rm GeV}^2$,
where reasonable experimental data are available, the
contributions of the massive $c$-quarks \cite{mass}
represent corrections contributing
about $3$-$4 \%$ to $\Delta(Q^2)$. Nuclear
effects of the iron target \cite{kulagin}
contribute about $1 \%$ to $\Delta(Q^2)$ 
[$\Delta_{\rm Fe}(Q^2) \approx 4 \cdot 10^{-3} {\rm GeV}^2/Q^2$].
The ``light-by-light'' contribution in the
GLS TPS (\ref{GLSTPS}) (it is $\sim\!a^3$) is
kept apart, it contributes to $\Delta(Q^2)$ less than
$0.5 \%$.

We work in the ${\overline {\rm MS}}$ renormalization scheme.
The renormalization scale $\mu$ was fixed according to
the principle of minimal sensitivity (PMS), obtaining
$\mu^2 \approx 3.3Q^2$. Experimentally extracted
values of GLS $\Delta(Q^2)$ at $Q^2 = 3.16 \ {\rm GeV}^2$
are given by the Fermilab CCFR
Collaboration \cite{Kim:1998ki}
\begin{equation}
\Delta(Q^2\!=\!3.16 {\rm GeV}^2) = 0.149 \pm 0.039 \ ,
\label{Deltaexp}
\end{equation}
where the nuclear corrections were subtracted out.
When we confront (\ref{Deltaexp}) with the theoretical GLS expression
[Eq.~(\ref{res1}) with the conformal mapping (\ref{ct}); 
plus $c$-quark corrections; plus ``light-by-light;''],
the obtained QCD $\alpha_s$ in ${\overline {\rm MS}}$ is
\begin{equation}
\alpha_s(Q) = 
0.305^{+0.132}_{-0.074} (\rm exp) \pm 0.006 (\rm th) \ , \
\alpha_s(M_Z) = 
0.1167^{+0.0128}_{-0.0118} (\rm exp) \pm 0.0008 (\rm th).
\label{res3GeV}
\end{equation}
Using the central CCFR data value of $\Delta(Q^2)$
at $Q^2\!=\!2.00 \ {\rm GeV}^2$, we obtain
almost the same central value: $\alpha_s(M_Z)= 0.1166$.
The values of $\alpha_s(M_Z)$ obtained are in agreement with
the present world average. The experimental uncertainties are
large. The theoretical uncertainties are dominated by
the ambiguity with respect to the change of the
renormalization scheme.

The term proportional to $\cot ( \pi \nu)$ in (\ref{res1}),
which eliminates the unphysical cut, was important
numerically as it decreases the value of $\Delta(Q^2)$
by about $10 \%$, thus increasing the extracted value
of $\alpha_s(M_Z)$ by about $0.003$. The CCFR Collaboration
\cite{Kim:1998ki}, by evaluating the NNLO TPS of $\Delta(Q^2)$,
corrected by a term $\propto\!1/Q^2$ as estimated by
QCD sum rules, obtained from the central values of the
data $\alpha_s(M_Z) \approx 0.113$-$0.114$. An earlier similar
(TPS-PMS) evaluation \cite{Chyla:1992cg} gave 
$\alpha_s(M_Z) \approx 0.115$.

\begin{acknowledgments}
The work of C.C. and G.C. was supported by DGIP (UTFSM)
and FONDECYT (Chile) Grant No.~1010094, respectively. The work of
T. L. and K.S.J. was supported in part by the BK21 Core Program.
\end{acknowledgments}


\begin{thebibliography}{10}

\bibitem{Mueller:vh}
A.~H.~Mueller,
Nucl.\ Phys.\ B {\bf 250}, 327 (1985).

\bibitem{Lee:2001ws}
T.~Lee,
Phys.\ Rev.\ D {\bf 66}, 034027 (2002)
[arXiv:hep-ph/0104306].

\bibitem{Contreras:2002kf}
C.~Contreras, G.~Cveti\v c, K.~S.~Jeong and T.~Lee,
arXiv:hep-ph/0203201, to appear in Phys. Rev. D.

\bibitem{GLZN}
S.~G.~Gorishny and S.~A.~Larin, Phys. Lett. B {\bf 172}, 109 (1986);
E.~B.~Zijlstra and W.~Van Neerven, Phys. Lett. B {\bf 297}, 377 (1992);
S.~A.~Larin and J.~A.~M. Vermaseren, Phys. Lett. B {\bf 259}, 345 (1991).

\bibitem{Shuryak:1981kj}
E.~V.~Shuryak and A.~I.~Vainshtein,
Nucl.\ Phys.\ B {\bf 199} (1982) 451;
Nucl.\ Phys.\ B {\bf 201} (1982) 141.

\bibitem{ct1}
G.~Cveti\v c and T.~Lee,
Phys.\ Rev.\ D {\bf 64} (2001) 014030;
G.~Cveti\v c, C.~Dib, T.~Lee and I.~Schmidt,
Phys.\ Rev.\ D {\bf 64} (2001) 093016;
G.~Cveti\v c, T.~Lee and I.~Schmidt,
Phys.\ Lett.\ B {\bf 520} (2001) 222;
K.S.~Jeong and T.~Lee, hep-ph/0204150.

\bibitem{ct2}
I.~Caprini and J.~Fischer,
Phys.\ Rev.\ D {\bf 60} (1999) 054014;
Phys.\ Rev.\ D {\bf 62} (2000) 054007;
Eur.\ Phys.\ J.\ C {\bf 24} (2002) 127;
J.~Fischer, J.~Chyla and I. Caprini,
arXiv:hep-ph/0206098.

\bibitem{KS}
A.~L.~Kataev and V.~V.~Starshenko, 
Mod. Phys. Lett. A {\bf 10}, 235 (1995).

\bibitem{mass}
M.~Buza, Y.~Matiounine, J.~Smith and W.~L.~van Neerven,
Nucl.\ Phys.\ B {\bf 485}, 420 (1997);
J.~Bl\"umlein and W.~L.~van Neerven,
Phys.\ Lett.\ B {\bf 450}, 417 (1999).

\bibitem{kulagin}
S.~A.~Kulagin, Nucl. Phys. {\bf A640}, 435 (1998).

\bibitem{Kim:1998ki}
J.~H.~Kim {\it et al.} (CCFR Collab.),
Phys.\ Rev.\ Lett.\  {\bf 81} (1998) 3595.

\bibitem{Chyla:1992cg}
J.~Chyla and A.~L.~Kataev,
Phys.\ Lett.\ B {\bf 297}, 385 (1992).

\end{thebibliography}
\end{document}